# Performance of the Near-infrared coronagraphic imager on Gemini-South

Mark Chun[*a], Doug Toomey[b], Zahed Wahhaj[a], Beth Biller[a], Etienne Artigau[c], Tom Hayward[c], Mike Liu[a], Laird Close[d], Markus Hartung[c], Francois Rigaut[c], Christ Ftaclas[a]

[a]Institute for Astronomy, University of Hawaii, 640 N. A'ohoku Place, Hilo, HI 96720;
[b]Mauna Kea Infra-Red LLC; [c]Gemini Observatory; [d]University of Arizona

**ABSTRACT**

We present the coronagraphic and adaptive optics performance of the Gemini-South Near-Infrared Coronagraphic Imager (NICI). NICI includes a dual-channel imager for simultaneous spectral difference imaging, a dedicated 85-element curvature adaptive optics system, and a built-in Lyot coronagraph. It is specifically designed to survey for and image large extra-solar gaseous planets on the Gemini Observatory 8-meter telescope in Chile. We present the on-sky performance of the individual subsystems along with the end-to-end contrast curve. These are compared to our model predictions for the adaptive optics system, the coronagraph, and the spectral difference imaging.

## 1. INTRODUCTION

The Gemini-South Near-Infrared Coronagraphic Imager (NICI) is a specialized dual-channel camera with a dedicated Lyot coronagraph and 85-element curvature adaptive optics system designed to search for and image large Jupiter planets around nearby stars by spectrally differencing two images taken in or next to strong near-infrared methane features found in the atmosphere of large Jovian-type planets. NICI is currently undergoing commissioning on the Gemini-South telescope and a 50-night science campaign is soon to begin. The commissioning has proceeded in phases first focusing on basic functionality and integration with the telescope subsystems, bright guide star AO performance, and bright guide star high-contrast imaging (e.g. coronagraphy, SDI, and ADI). We present here the results of these commissioning activities.

The implementation of NICI largely follows unchanged from its origin design described by Toomey and Ftaclas[1] (2003). A number of recent instruments have been deployed for similar science objectives (CFHT/Trident[3], NICI+Altair[4], VLT NACO/SDI[5]) but NICI represents the first of such systems that takes a system-wide approach to everything after the telescope focal plane for this specific science objective. The NICI design integrates the three major subsystems (AO relay, coronagraph, and dual-channel camera) into a single facility-class instrument with the design philosophy that it be limited only by the residual atmospheric wavefront and scattering. As such, all effort was made to minimize optical scattering within the instrument and to minimize non-common path aberrations. The reader is referred to Toomey and Ftaclas (2003) for details but we highlight here a few of the features of the system that make its use or performance unique.

Figure 1 below shows the major subsystems of NICI. The entrance focus of the NICI is the f/16 Cassegrain focus of the Gemini-South 8-meter telescope. The AO relay is a standard single pupil-conjugated DM design with two exceptions. First, the adaptive optics system uses a curvature sensor/mirror with 85 correcting elements and, second, the dichroic that splits the light for the wavefront sensor and the science camera reflects near-infrared light to the science channel. This approach with the dichroic minimizes ghosting and chromatic aberrations with an all-reflective science optical train up to the transmissive focal plane occulting mask of the coronagraph. Within the wavefront sensor channel a field-steering mirror (at a pupil position) selects the guide star from the 18-arcsecond diameter field of view and a membrane mirror (at a focal plane) adjusts the optical gain of the curvature sensor. In the science channel the AO-corrected wavefront is imaged on to one of several focal plane masks held in a rotating mechanism just outside the camera dewar window. The masks are transmissive and have flat-top Gaussian-shaped transmission functions[2] (Ftaclas et al. 2008 these proceedings). Notably, the focal plane masks are not fully opaque at the center but were designed to provide partial transmission for astrometry and registration of images in the data reductions. In principle photometry could also be reconstructed from the final images though this must take into account the focal plane transmission function, the point spread function, and the stars location on the focal plane mask.

*mchun@ifa.hawaii.edu; phone 1 808 932-2317; fax 1 808 933-0737

Immediately after the focal plane mask, the science beam passes into the dewar through a window. The camera optical design is also fairly standard using diamond-turned off-axis parabolas in a 2:1 focal length enlarger. The pupil plane mask of the coronagraph is implemented in two stages first as a spider-mask and inner Lyot stop (hanving an equivalent central obscuration of 27%) followed by a separate outer pupil mask. The outer Lyot stops are placed in a wheel and a variety of fixed pupil stops can be selected from 80%-120% pupil diameter stops. The spider masks (30 times over sized) and the inner Lyot stop are combined into a single rotating mechanism that is always in the optical path. This greatly simplifies the implementation but an important, undesirable, ramification of this implementation is that the coronagraph can *never* be fully retracted from the optical path. The outer Lyot stop can be removed but the inner Lyot stop is always in place.

After the coronagraph the beam is split between the two science channels. This beamsplitter wheel has multiple positions and allows the beam to be split in a number of ways. At the moment the beam split is made using a simple 50/50 beam splitter. A dichroic is highly desirable here but given the narrow spacing between the H-band methane filters, a suitable dichroic design was not achieved. The two science channels differ after the beam split and each has its own filter wheel, focusing mirror, detector array (1024 x 1024 InSb), and array controller.

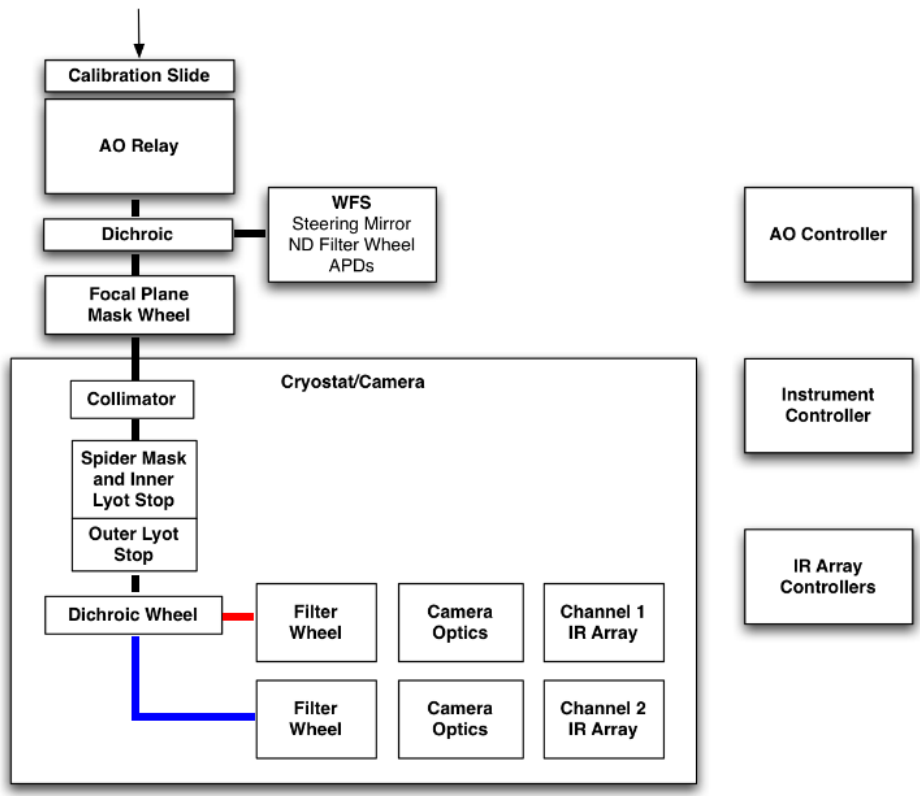

*Illustration 1: Block diagram of the major subsystems of NICI*

Before we discuss the performance of the NICI subsystems we note that the optimal use of NICI for high-contrast imaging is still being defined as commissioning proceeds. To date the instrument's default mode of observing uses the AOS on-axis, the coronagraph with a flat-topped Gaussian focal plane mask and a 90% hard-edged pupil-plane mask, spectral-difference imaging (SDI) with 4% filters around the 1.6 micron methane feature, and angular-difference imaging (ADI). While some tests have been done with other SDI filters and without ADI, these other modes are predominantly untested. In addition, the optimization of the AOS and coronagraph have only just begun. For example, the AOS has been run at a single extra-focal distance and only one combination of the the coronagraph occulting mask and Lyot stop has been used. Nonetheless, as will be shown in Sections 3 and 4, NICI is already obtaining excellent contrast limits.

While it is clear that the combination of ADI and SDI (ASDI) is very powerful, the ADI mode of observing has important penalties. ADI allows the science field to rotate at the science detector plane by observing with the Gemini Cassegrain rotator off. This field rotation introduces a field dependent blurring that reduces the Strehl and hence the planet detection sensitivity at large radii. This effect can be limited by setting a maximum exposure time per image and/ or by observing in positions of the sky when the field rotates slowly, however, these must be traded against the increased contribution from read noise and the degraded performance of the adaptive optics system due to the larger observing zenith distances (Biller et al. 2008).

## 2. CAMERA PERFORMANCE

The details of the NICI camera design and layout are provided in Toomey and Ftaclas (2003) and are not repeated here. Mechanical flexure is very low between the AO WFS, focal plane mask, and the detectors. There is less than 30 mas differential flexure between the two channels and less than 0.2 arcsecond (on the focal plane) flexure between the AO WFS and the focal plane mask over *all* elevation angles. The flexure is well behaved and it and the differential refraction are modeled into the Gemini instrument control software to account for these effects. The detector array controller performance meets the design requirements although detector read noise is the current limitation in achieving high contrasts at large separations from stellar images. The read noise per double correlated read is 60 e- but the controller can read non-destructively (NDR) up to 8 times to bring the read noise down to about 24 e-. The array controller is being reprogrammed to remove this limitation.

One significant change from the original camera design are the science channel methane filters. The original 1.6-micron methane filter set are 1% wide but new 4%-wide filters were chosen after a detailed simulation of the end-to-end propagation of light through NICI that included residual wavefront errors from the AOS, the coronagraph, and chromatic variations across the filter bandpass (see Section 4).

## 3. PERFORMANCE OF THE ADAPTIVE OPTICS SYSTEM

The adaptive optics system in NICI consists of an 85-element curvature wavefront sensor and deformable mirror. The AO control electronics, software, deformable mirror, and wavefront sensor were designed and built by the Institute for Astronomy at the University of Hawaii. Integration and initial commissioning of the NICI AOS was done with a deformable mirror from CILAS but this was later replaced by deformable mirror by the University of Hawaii with a smaller minimum radius of curvature and higher resonance frequency (Ftaclas et al. 2008).

### 3.1 Performance predictions and lab characterization

The performance of the adaptive optics system was simulated using the simul IDL package written by Francois Rigaut. The simulation uses a median optical turbulence and wind profiles for Cerro Pachon (Vernin et al. 2000) where the Gemini telescope is located. Telescope static aberrations are assumed to be limited to low-order aberrations and corrected by the telescope's active optics. The deformable mirror's minimum radius of curvature for the 400V maximum voltage provided by the deformable mirror electronics was matched to the UH DM's minimum radius of curvature of 13.1 meters surface for the central actuator. The sampling frequency of the simulated AO system was 1kHz and it should be noted that the simulation also samples wavefronts (instantaneous) at this same frequency. Finally, non-common path aberrations within the adaptive optics system and camera are modeled as a static wavefront error after the AO correction. Here we used a value of 140 nm RMS based on the static Strehl=0.75 at 1.6 microns achieved when the AOS was closed loop on the internal calibration source during acceptance testing of the instrument. This value surpasses our error budget for residual optical figure errors and alignment errors for the combination of the camera and adaptive optics system.

Items that are not included in the simulation are the long-term atmospheric image motion, telescope jitter, and the elongation of the extra-pupil images within the wavefront sensor inherent in the optical design of the wavefront sensor. For the performance of the adaptive optics system we limited the equivalent total integration times to a few seconds. For the coronagraphic simulations, long-exposure images were simulated by cycling the phase screen in the simulation to effectively average over a larger number of realizations so while these also totaled a similar length of equivalent integration time, the temporal power spectrum of the phase aberrations are effectively truncated at low frequencies much less than the 0dB bandwidth of the system.

Figure 2 below shows the predicted performance of the adaptive optics system. Both plots include an static wavefront error of 140nm RMS and the simulations were all run at a wavelength of 1.6 microns. The delivered Strehl is expected to be better than 35% at 1.6 microns for a bright guide star under median seeing conditions.

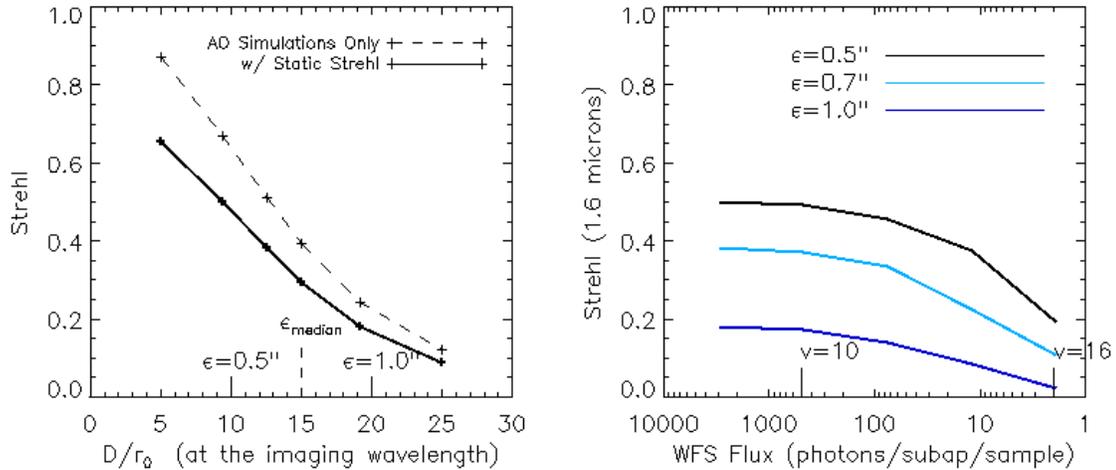

*Illustration 2: NICI AO Performance Simulations. Strehl is plotted versus the seeing ($D/r_0$) and the guide star brightness. In addition, the AO-only (excluding non-common path errors) is shown in the plot on the left as a dashed line. The guide star brightness is plotted in terms of the number of photons per subaperture per WFS sample. For reference, typical seeing values at 0.5 microns are labeled in the plot of the left at their respective $D/r_0$ for 1.6 micron imaging and in the plot on the right a guide star with magnitude $V=10$ corresponds to about 500 photons/sub/sample.*

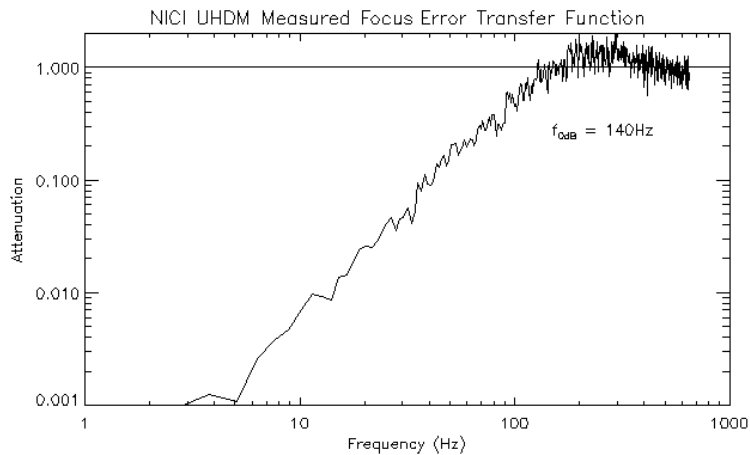

*Illustration 3: Measured control servo bandwidth on the Zernike focus mode for the NICI AOS using University of Hawaii deformable mirror measured by closing the loop on photon noise in the wavefront sensor and projecting the noise onto the Zernike focus mode. The sampling frequency of the system is 1.3kHz and the loop gain is $g=0.5$. The 0dB bandwidth is well in excess of the 100Hz design goal. The input noise spectrum is white noise and the curve is normalized to unity at large frequencies.*

During integration of the University of Hawaii DM into NICI, we measured the control bandwidth of the system by closing the control servo loop on photon noise in the wavefront sensor and constructing the equivalent Zernike focus error transfer function. The 0dB bandwidth is between 130-140 Hz. This represents an increase of about a factor of five over the initial DM deployed in the system since importantly the UHDM does not introduce any dynamic (vibration) errors at higher servo gains. The UHDM has hysteresis (~20%) but this does not impact the dynamic bandwidth of the system.

## 3.2 On-sky Characteristics

To date the commissioning of NICI has focused on bright star performance in the full AO + coronagraph + ASDI. As such, the adaptive optics system performance reported here is mostly limited to guide stars V~11 magnitude and brighter. In addition to the usual commissioning hurdles, two noteworthy surprises included large oscillations of the telescope top-end (up to 100 milliarcseconds peak-to-peak) found at frequencies in excess of 15Hz and an apparent change in focus between the science camera and the adaptive optics system. The telescope top-end oscillations were resolved during a reassembly of the telescope secondary assembly while the focus offset between the camera and AOS was traced to an error in the DM control electronics coupled with the hysteresis in the UHDM and the hidden mode of the system.

The bright guide star performance of the AO system is illustrated in Figure 4 below. The delivered Strehl, as measured on the two detector focal planes, are from images taken during the last two commissioning runs of stars off the focal plane mask (either images with the focal plane mask removed or companion/background objects within the field of view of coronagraphic images). The Strehl is calculated by estimating the peak and flux of the star image and calculating a theoretical point-spread function with an appropriate wavelength, pixel sampling, and coronagraphic pupil mask. Seeing is estimated from AO telemetry data by taking the influence functions of the deformable mirror, with a minimum radius of curvature of 13.1 meters surface for the central actuator, and reconstructing the corrected Zernike coefficients from the DM voltages. $D/r_0$ is then derived from the variance of the higher-order Zernike terms (focus to Zernike 36) and the expected variances (Noll 1976). A cross check of the seeing reconstructed in this way agrees with the seeing monitor outside the Gemini-South enclosure but a detailed cross comparison between the two has not been done.

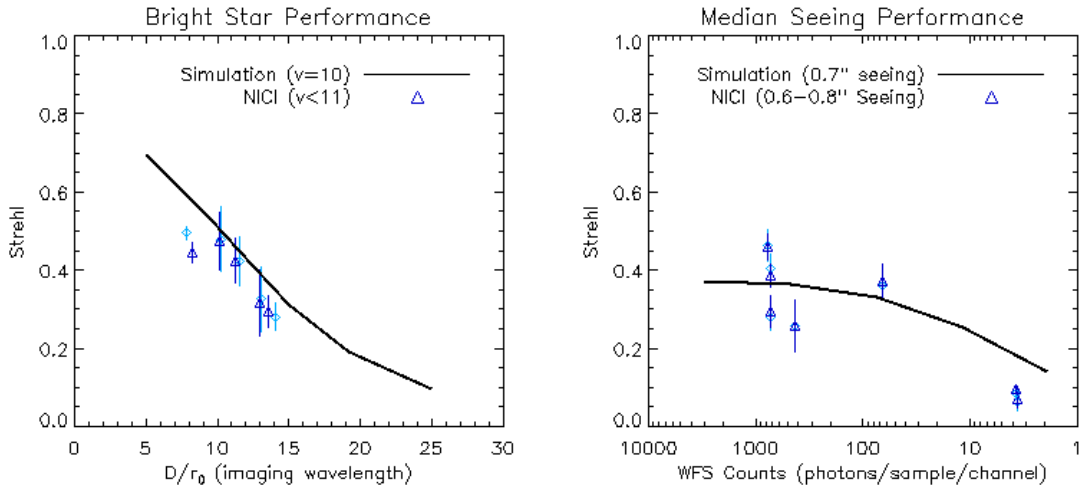

*Illustration 4: NICI's adaptive optics delivered performance compared with simulations. The Strehl ratios are measured on the detector focal plane of long-exposure images and include all non-common path errors. The simulation curves account for the measured static Strehl of 0.8 at 1.6 microns as measured on NICI's internal calibration source. The $D/r_0$ are measured from the AO telemetry data (see text). The WFS counts are per channel per wavefront sample and NICI nominally samples the wavefront curvature at 1.3kHz.*

The bright star performance already agrees well with the Strehl ratios predicted from the simulations. The plots represent all of the data from the last two commissioning runs for which there is an image with a linear stellar companion image off the focal plane mask and an AO telemetry data set taken within 5 minutes of the image. The seeing during these runs was better than the median seeing for the site. The site's median seeing (0.7") corresponds to a

$D/r_0$=15 at 1.6 microns. For bright guide stars ($V_{gs}$ = 11 or brighter) the measured Strehl ratios agree well with the predicted performance curve. The performance appears to saturate at $SR_{1.6um} \sim 50\%$ under very good seeing conditions but we note that the curvature wavefront sensor has been used at only one optical gain (extra-focal distance = 0.4m) during all of the on-sky commissioning. At these seeing values the optimal extra-focal distance determined from the simulations is smaller and from the simulations this difference would account for most of this difference. We have only begun to use the system on fainter guide stars and the performance of the system at the moment is worse than expected for a V=14 magnitude star.

Racine (2006) found that for most implemented adaptive optics systems the actual delivered residual wavefront phase variance normalized by $(D/r_0)^{5/3}$ is simply related to the number of correcting elements and the type of adaptive optics system deployed. The NICI AOS with 85 sensing/correcting curvature elements follows the trend Racine found for lower order curvature systems though. Figure 5 below shows Figure 1 from Racine (2006) and where the NICI AOS "bright guide star" performance falls on his distribution. Given that the loop gains (servo and optical) have yet to be tuned the results are very encouraging.

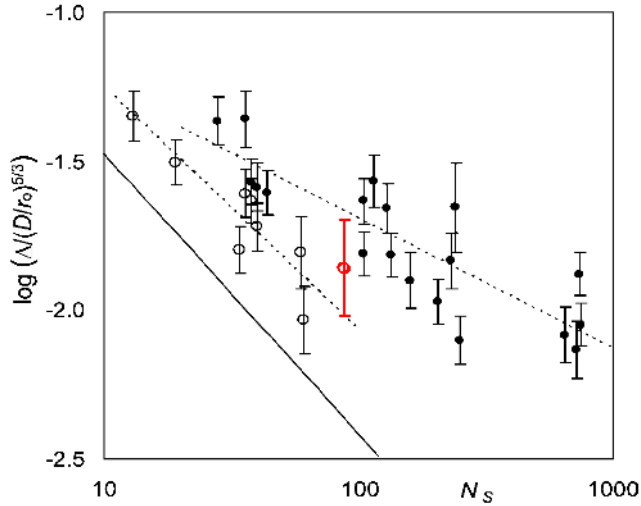

*Illustration 5: Figure 1 from Racine (2006) showing the relationship between the delivered residual phase error and the number of actuators in the system. Open circles are curvature systems while solid dots signify Shack-Hartmann systems. The NICI AOS is the red open circle at $N_S$=85. The length of the bar for this point signifies the range in the residual phase variance over the course of the last two NICI commissioning runs.*

The current focus of work on AO is to automate the servo and optical gains for better performance at all guide star brightnesses and to integrate the AO telemetry data with the image data for seeing estimates and PSF reconstruction. These will be an important inputs to the science campaign's final contrast curves both to quantify the achieved contrasts under different seeing conditions but also as an estimate of the underlying non-coronagraphic stellar point-spread function.

## 4. CORONAGRAPHIC PERFORMANCE

### 4.1 Simulations

The end-to-end simulation of the delivered NICI images were done in two stages. First, the adaptive optics systems corrected images were generated using a Monte-Carlo simulation package (simul) originally written by Francois Rigaut. The output instantaneous wavefronts were then propagated through a coronagraph code (based on code by Christ Ftaclas) both on (star) and off (planet) the focal plane mask. These generated point spread functions (both coronagraphic and non-coronagraphic) were generated at multiple wavelengths across the H-band and K-bands. The output data cube of images were then combined (assuming a flat spectrum object outside the methane absorption band), and then scaled to various brightnesses and radial distances from the central star image (coronagraphic). These final images were then run through a data pipeline to produce an SDI image and an associated noise profile.

The simulation of the AO corrected images is discussed in Section 3. In the coronagraphic simulation code we propagate the AO corrected complex wavefronts to the first focus. Here the image amplitude is modified by the occulting focal plane mask and propagated to the next pupil plane where the Lyot stop is applied (again in amplitude). Finally the image is propagated back to the second focus where we integrate the final image. This is done at multiple wavelengths to produce a final image cube. In addition, complex wavefronts are propagated with and without the occulting focal plane mask with the latter providing a PSF for the planet image. We ignore any angular decorrelation of the wavefront between the star and the planet and assume that the planet image will be close to the stellar image.

Several important factors are not included in these simulations. Due to the computational demand of the Monte-Carlo approach, the total integration time we could simulate was limited. To compensate for this, we rotated and randomly shifted the phase screens after each 0.1-seconds of integration. The AO control servo was allowed to converge before the integration was continued. In this way, the final image consists of the combination of many short exposures from over 300 independent realizations of the atmosphere. Similarly, due to the computational demand, the simulations do not include simulation of the ADI where correlations of the speckle pattern (AO corrected complex wavefront) over many seconds to minutes is needed to properly simulate the effect. Finally, low-frequency misalignment of the star on the focal plane mask (due to for example flexure, atmospheric differential refraction between the AO WFS and the science channel) is not included. This in practice is an important requirement and we note that an advantage of the partially transmissive occulting mask plus a repeatable focal plane mask mechanism enables a simple feed-back procedure to maintain this alignment.

We characterize the end-to-end performance of NICI by the final contrast achieved in the methane H-band filters. Figure 6 below shows the predicted contrast curve for NICI. The contrast curve is a strong function of the seeing and guide star brightness so the plotted points represent excellent seeing at the site and a relatively bright guide star.

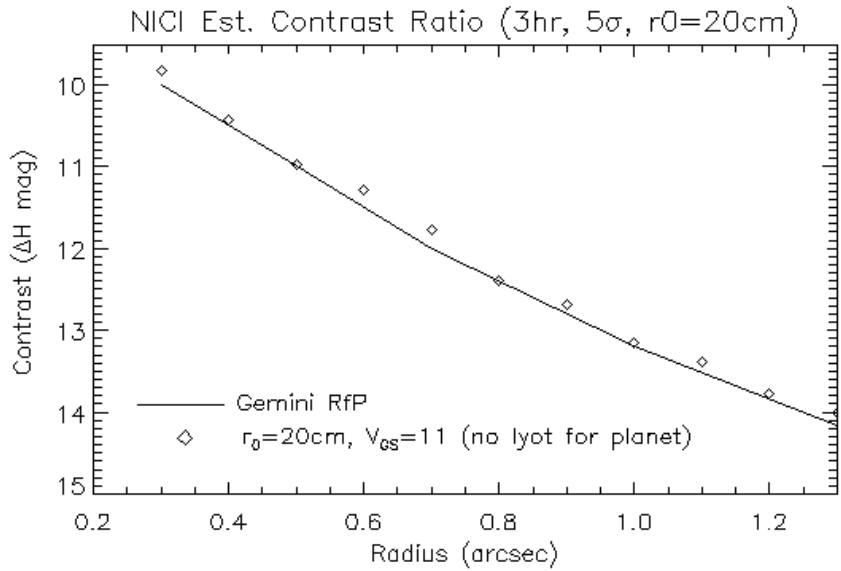

*Illustration 6: NICI contrast prediction under better than average seeing and a bright guide star. The solid line represents the assumed contrast curve issued by the Gemini Request for Proposals for the 50-night NICI science campaign while the diamonds represent the values from our simulations. Note that these simulations assumed a deformable mirror with a minimum radius of curvature of $R_{min}$=17.5m. The UHDM in NICI achieves a 13.1m minimum radius of curvature but under good seeing conditions the performance should not be limited by the minimum radius of curvature so the performance should similar.*

### 4.2 Performance

During commissioning the main mode of observing has been a combination of the coronagraph + ADI/SDI (ASDI) on reasonably bright guide stars (V~11th magnitude or brighter). Commissioning targets were selected from previous surveys with known background companions with measured photometry.

Intimately tied to the delivered performance of the system is the data reduction and analysis pipelines. As part of the preparations for the science campaign one of us (Biller) has adapted the University of Arizona SDI pipeline (Biller et al 2007) to NICI and Wahhaj and Artigau have written two new pipelines. The reader is referred to Artigau et al. (2008) and Biller et al. (2008) where more details on the pipelines can be found.

In summary, the three pipelines all follow these basic steps: (1) normal image reductions (flat fielded, dark subtraction, and bad pixel removal), (2) high-pass filter the image (for example removing azimuthal average/median profile), (3) generating a 'quasi-static' PSF for removal taking advantage of the ADI observing, (4) registration and subtracting the two filters to produce the SDI images, and (5) finally rotating all of the ASDI images to a common field position angle and combining to create a final science image frame. The precise order and sequence of the above steps differs slightly in the three pipelines as does the particular algorithms deployed in each step (e.g. generating the reference PSF). However, we constructed a fake planet data set using one of the commissioning data sets and all three pipelines detect the fake planets with similar signal-to-noise.

Figure 7 below shows the contrast curve obtained by one of the pipelines (Wahhaj) on three of the fields observed during commissioning. The three data sets represented by solid lines were taken under median to better than median seeing conditions with the Cassegrain rotator off (ADI mode). The integration times were only about 30 minutes and were made on stars with brightnesses between v=8 and v=11. The curves illustrate excellent contrasts can be achieved from the occulting mask edge (0.32as radius) to 1 arcsecond from the star. Within this region the contrast is limited by the speckle subtraction. The contrast past 1 arcsecond should start to be dominated by photon noise limited but for now is limited by the read noise of the detector and the limited number of non-destructive reads (NDR) currently possible with the array controller. Note that the observations of the faintest of the three targets (TWA7) had only 4 NDR while the other two fields used the maximum 8 NDR. This highlights ADI's strong requirement on minimizing detector noise. The plot on the right in Figure 7 shows the contrast curves scaled to a 2-hour integration. The data to date shows that the contrast increases, as expected, as the square root of the exposure time. For comparison the contrast curves specified in the Gemini Science Campaign Request for Proposals (RfP) and obtained during the Gemini Deep Planet Survey (Lafrenière et al 2007) are shown. The additional magnitude of contrast is likely due to the simultaneous two-channel imaging possible with NICI (Artigau et al. 2008).

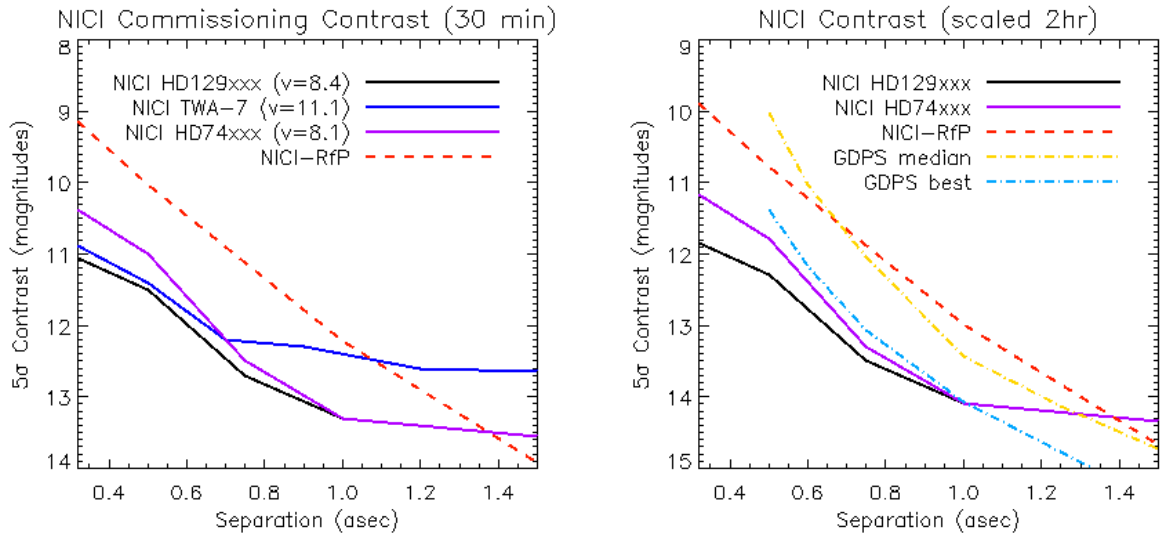

*Illustration 7: NICI contrast curves achieved commissioning. The curves shown are from the Wahhaj pipeline but all three data pipelines obtain similar contrasts. The contrast curves on the left shows the achieved contrasts for the actual integration times (approximately 30 minutes) as well as the NICI simulation/RfP contrast curve scaled to 30 minutes. The V-band brightnesses of the stars are listed in the legend. Note that TWA-7 is considerably fainter than the other two targets and, in addition, the number of non-destructive reads was less than that for the other two fields. The figure on the right shows the contrast curve obtained on the two brighter commissioning targets scaled by the square root of the integration time to 2 hours. For reference the NICI-RfP curve and two contrast curves from the GDPS are also shown scaled to this same total integration time.*

## 5. CONCLUSIONS

The performance of the Gemini-South NICI instrument is progressing well. AO performance matches predictions and the combination of ADI and SDI provides contrasts in excess of the original predictions. At the moment the contrast at separations less 1" are limited by the speckle differencing while at larger radii we are limited by detector read noise. Work is focusing on minimizing this by increasing the number of nondestructive reads the detector array controller performs. NICI's integrated approach with AO, coronagraph, ADI, and SDI that make it stand out from past instruments with similar science objectives and current performance bears out the advantages their combined use.

A 50-night science campaign with NICI will begin at the end of the instrument commissioning targeting nearby young stars (t$\leq$500Myr) to search for young hot Jupiter-mass planets. The phase space of the campaign complements existing radial velocity surveys by looking for planets at larger separations from their stars.